\documentclass{article}%
\usepackage{amsfonts}
\usepackage{amsmath}%
\usepackage{amssymb}%
\usepackage{graphicx}
\providecommand{\U}[1]{\protect\rule{.1in}{.1in}}
\begin{document}

\title{A visual representation of the properties of pre- and post- selected entangled systems}
\author{Yakir Aharonov$^{a,b}$ and Tomer Shushi$^{c}$\\$^{a}$Schmid College of Science, Chapman University, Orange\\CA 92866, USA\\$^{b}$Raymond and Beverly Sackler School of Physics\\and Astronomy Tel-Aviv University, Tel-Aviv\\69978, Israel\\$^{c}$Center for Quantum Science and Technology,\\Department of Business Administration,\\Guilford Glazer Faculty of Business and Management,\\Ben-Gurion University of the Negev, Beer-Sheva, Israel}
\maketitle

\begin{abstract}
We introduce a visual representation for generating entangled-based quantum
effects under pre- and post- selected states that allows us to reveal
equivalence between seemingly different quantum effects. We show how to
realize entangled quantum systems of an arbitrary number of qubits from a
single or pre-specified number of physical particles. We then show that a
variation of the quantum Cheshire cat experiment and Hardy's paradox are
equivalent and propose a class of experiments that generalizes both
experiments. We show that the weak values of the products of projection
operators allow us to get the weak value of each projection operator, implying
that the weak value of the product of projection operators includes the entire
information about the weak values in the system. In nature, interactions can
only be acted between a pair of particles. We show how to realize quantum
systems of multiwise interacted qubits, i.e., interactions that come in groups
of n%
%TCIMACRO{\TEXTsymbol{>}}%
%BeginExpansion
$>$%
%EndExpansion
2 qubits. In this way, we are able to propose unique quantum systems that
consist of interacted groups of entangled states. The proposed framework opens
the door toward a new way to explore quantum systems of entangled particles
and quantum phenomena that emerge from such a general setting.

\textit{Keywords: entanglement, Hardy's paradox, multiwise interactions,
quantum computation, quantum Cheshire cat experiment, qubits}

\end{abstract}

\bigskip\newpage

\section{Introduction}

In quantum mechanics, the properties of particles share the same mathematical
description as quantum states in some Hilbert space. Thus, for instance, a
Schr\"{o}dinger equation of $N$ particles in one spatial dimension is
mathematically equivalent to a single particle in $N$\ dimensions. In quantum
computing, there are different ways to derive qubits followed by this fact,
for instance, creating physical qubits from photons, also known as Photonic
Quantum Computing, or creating qubits using electrons in quantum dots, known
as Quantum dot qubits [1-5]. We consider a pair of qubits $\left(
I,II\right)  $ and define a scheme of expected projection operators where each
of the components is%
\begin{equation}
a_{ij}:=\left\langle \psi\left\vert \Pi_{i}^{I}\Pi_{j}^{II}\right\vert
\psi\right\rangle ,\text{ }i,j=1,2,\label{4a}%
\end{equation}
for the projection operators $\Pi_{i}^{I/II}=\left\vert i\right\rangle
\left\langle i\right\vert ,i=0,1,$ and so $a_{ij}$ are the components of
the\ scheme given in Figure 1.

{\includegraphics[
height=2.2in,
width=2.3in
]{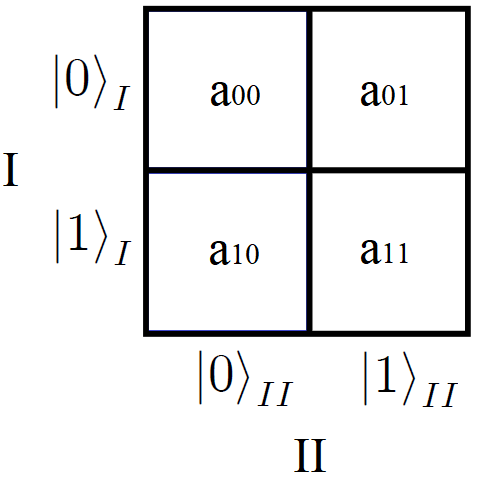}}

Figure 1. The scheme of the realized pair of particles based on the expected
values of $\Pi_{i}^{I}\Pi_{j}^{II},i,j=1,2.$

\bigskip

As a special case of Figure 1, consider quantum systems that follow the
components
\begin{equation}
a_{12}=a_{21}=1/2,\text{ }a_{ii}=0,\text{ }i=0,1.\label{ampl12}%
\end{equation}
Then, the system leads to the Bell states%
\begin{equation}
\left\vert \Psi^{\pm}\right\rangle =\frac{\left\vert 0\right\rangle
_{I}\left\vert 1\right\rangle _{II}\pm\left\vert 1\right\rangle _{I}\left\vert
0\right\rangle _{II}}{\sqrt{2}},\label{Bell1}%
\end{equation}
and%
\begin{equation}
\left\vert \Phi^{\pm}\right\rangle =\frac{\left\vert 0\right\rangle
_{I}\left\vert 0\right\rangle _{II}\pm\left\vert 1\right\rangle _{I}\left\vert
1\right\rangle _{II}}{\sqrt{2}}.\label{Bell2}%
\end{equation}
By adding another physical particle to the given system, we obtain a system of
$4$ realized qubits followed by the system of operators%
\begin{equation}%
%TCIMACRO{\QATOPD{\{}{.}{\left(  \left(  \sigma_{Z}^{I},\sigma_{X}^{I}%
%,\sigma_{Y}^{I}\right)  ,\left(  \sigma_{Z}^{II},\sigma_{X}^{II},\sigma
%_{Y}^{II}\right)  \right)  }{\left(  \left(  \sigma_{Z}^{III},\sigma_{X}%
%^{III},\sigma_{Y}^{III}\right)  ,\left(  \sigma_{Z}^{IIII},\sigma_{X}%
%^{IIII},\sigma_{Y}^{IIII}\right)  \right)  }}%
%BeginExpansion
\genfrac{\{}{.}{0pt}{}{\left(  \left(  \sigma_{Z}^{I},\sigma_{X}^{I}%
,\sigma_{Y}^{I}\right)  ,\left(  \sigma_{Z}^{II},\sigma_{X}^{II},\sigma
_{Y}^{II}\right)  \right)  }{\left(  \left(  \sigma_{Z}^{III},\sigma_{X}%
^{III},\sigma_{Y}^{III}\right)  ,\left(  \sigma_{Z}^{IIII},\sigma_{X}%
^{IIII},\sigma_{Y}^{IIII}\right)  \right)  }%
%EndExpansion
.\label{6}%
\end{equation}
This corresponds to the three-dimensional tensor $2\times2\times2$, and let us
consider, for simplicity, equal-valued tensor components%
\[
a_{ijk}=1/3,\text{ \thinspace}i\neq j\neq k,i,j,k=0,1,\text{ and }%
a_{ijk}=0\text{ otherwise.}%
\]
Using only the first three triplets, we can realize a GHZ state that is
established as the eigenstate of the considered operators%
\begin{equation}
\sigma_{Y}^{I}\sigma_{Y}^{II}\sigma_{X}^{III},\,\sigma_{Y}^{I}\sigma_{X}%
^{II}\sigma_{Y}^{III},\text{ }\sigma_{X}^{I}\sigma_{Y}^{II}\sigma_{Y}%
^{III},\text{ }\sigma_{X}^{I}\sigma_{X}^{II}\sigma_{X}^{III},\label{7}%
\end{equation}
with the GHZ (eigen-)state%
\begin{equation}
\frac{1}{\sqrt{2}}\left(  \left\vert 0\right\rangle _{I}\otimes\left\vert
0\right\rangle _{II}\otimes\left\vert 0\right\rangle _{III}+\left\vert
1\right\rangle _{I}\otimes\left\vert 1\right\rangle _{II}\otimes\left\vert
1\right\rangle _{III}\right) \label{8}%
\end{equation}
The GHZ state can also be realized using only a single physical particle but
with a cube instead of a two-dimensional square, allowing us to consider eight
equal sub-cubes. For such a system,\ the gate is then defined through three
triplets $\left(  \sigma_{Z}^{I},\sigma_{X}^{I},\sigma_{Y}^{I}\right)
,\left(  \sigma_{Z}^{II},\sigma_{X}^{II},\sigma_{Y}^{II}\right)  ,$ and
$\left(  \sigma_{Z}^{III},\sigma_{X}^{III},\sigma_{Y}^{III}\right)  .$ We can
further extend it into\ $n-$dimensional hypercube to obtain the entangled
state of a system of qubits
\begin{equation}
\left\vert \Psi\right\rangle =%
%TCIMACRO{\dsum \limits_{j=1}^{2^{n}}}%
%BeginExpansion
{\displaystyle\sum\limits_{j=1}^{2^{n}}}
%EndExpansion
\alpha_{j}\left\vert j_{1}\right\rangle _{I}\otimes\left\vert j_{2}%
\right\rangle _{II}\otimes...\otimes\left\vert j_{n}\right\rangle
_{\underset{n}{\underbrace{I...I}}}.\label{9}%
\end{equation}
We have thus established an equivalency between $k$ physical particles and
$N>k$ particles that form realized entangled states, even in the case of
single physical particles.

In the following, we present a visual representation of quantum systems of
entangled particles that allows us to obtain fundamental structures of
entangled particles to both obtain new experimental setups of entangled
systems and serve as a tool for detecting relations between seemingly
different quantum effects.

\section{\bigskip Results}

The two-state-vector formalism (TSVF) is a time-symmetric description of
quantum mechanics, followed by two boundary conditions to quantum systems,
namely, an initial state $\left\vert \psi\right\rangle $ and a final state
$\left\vert \phi\right\rangle ,$ which are also known as the pre- and post-
selected states of the system [8-12]. Using this formalism, one can establish
a unique way to measure quantum systems without the collapse of the
wavefunction, known as \textit{weak measurements}. Unlike strong measurements,
which collapse the wavefunction of a quantum system into a definite state,
weak measurements provide partial information about the system's state while
keeping the system in a superposition of possible outcomes. This is achieved
by weakly coupling the measuring device to the quantum system, resulting in a
small shift in the measurement apparatus without fully collapsing the system's
state. A \textit{weak value} is the outcome of this intermediate, weak
measurement, and it can sometimes take surprising results [10-14]. Unlike
standard measurement outcomes, weak values can be complex numbers or can lie
outside the range of eigenvalues of the observable being measured. This
phenomenon arises due to the non-trivial interplay between the pre-selected
and post-selected states. The weak value of an observable $A$ is defined by%
\begin{equation}
\left\langle A\right\rangle _{w}=\frac{\left\langle \phi\left\vert
A\right\vert \psi\right\rangle }{\left\langle \phi|\psi\right\rangle
}\label{Aw1}%
\end{equation}
where the pre- and post- selected states are not orthogonal, i.e.,
$\left\langle \phi|\psi\right\rangle \neq0.$

We extend the proposed framework into the weak quantum reality, which allows
us to capture the two boundary conditions of the system. This is done by
proposing the following weak values of projection operators%
\begin{equation}
a_{ij}=\frac{\left\langle \phi\left\vert \Pi_{i}^{I}\Pi_{j}^{II}\right\vert
\psi\right\rangle }{\left\langle \phi|\psi\right\rangle }\label{awij1}%
\end{equation}
for the components of the proposed scheme. The weak values of the product of
projection operators (\ref{awij1}) allow us to get the weak value of each
projection operator, i.e., $\sum_{i=1}a_{ij}=\left\langle \Pi_{j}%
^{II}\right\rangle _{w}$ and $\sum_{j=1}a_{ij}=\left\langle \Pi_{i}%
^{I}\right\rangle _{w}$. This property indicates that the product of
projection operators contains complete information regarding the weak value of
each projection operator within the system.

Let us now review the quantum Cheshire cat experiment and Hardy's paradox and
then show that they have, in fact, the same structure detected using the
proposed framework (\ref{awij1}).

We start with the quantum Cheshire cat experiment [6]. Suppose we have an
electron that can be in two boxes, denoted by $\left\vert L\right\rangle $
and\ $\left\vert R\right\rangle $ for the left and right boxes, respectively,
and assume that the electron is polarized along the $x-$axis. We then prepare
pre- and post- selection states of the electron in the form%
\begin{equation}
\left\vert \psi\right\rangle =\frac{1}{\sqrt{3}}\left(  \left\vert
L\right\rangle \left\vert \uparrow_{z}\right\rangle +\left\vert R\right\rangle
\left(  \left\vert \uparrow_{z}\right\rangle +\left\vert \downarrow
_{z}\right\rangle \right)  \right)  ,\label{preqc1}%
\end{equation}
and post-select the state $\left\vert \phi\right\rangle $ in the form%
\begin{equation}
\left\vert \phi\right\rangle =\frac{1}{\sqrt{3}}\left(  \left\vert
L\right\rangle \left\vert \uparrow_{z}\right\rangle -\left\vert R\right\rangle
\left(  \left\vert \uparrow_{z}\right\rangle -\left\vert \downarrow
_{z}\right\rangle \right)  \right)  .\label{preqc2}%
\end{equation}

Then, the position of the particle is given by the projection operator
$\Pi_{L/R}=\left\vert L/R\right\rangle \left\langle L/R\right\vert ,$ and the
spin is given by~$\Pi_{\uparrow_{z}/\downarrow_{z}}$. We can now establish the
quantum Cheshire cat effect. Under the pre- and post- selected states
(\ref{preqc1}) and (\ref{preqc2}), the weak value of position tells us that
the electron is on the left side while the spin along the $z-$axis is on the
right side, since%
\begin{align}
\left\langle \Pi_{L}\Pi_{\uparrow_{z}}\right\rangle _{w}  & =1,\left\langle
\Pi_{L}\Pi_{\downarrow_{z}}\right\rangle _{w}=0,\label{qc1}\\
\text{ }\left\langle \Pi_{R}\Pi_{\uparrow_{z}}\right\rangle _{w}  &
=-1,\left\langle \Pi_{R}\Pi_{\downarrow_{z}}\right\rangle _{w}=1,
\end{align}
implying that%
\begin{align}
\left\langle \Pi_{L}\right\rangle _{w}  & =\left\langle \Pi_{L}\Pi
_{\uparrow_{z}}\right\rangle _{w}+\left\langle \Pi_{L}\Pi_{\downarrow_{z}%
}\right\rangle _{w}=1,\text{ }\left\langle \Pi_{R}\right\rangle _{w}%
=0\label{qc2}\\
\left\langle \Pi_{\downarrow}\right\rangle _{w}  & =\left\langle \Pi_{L}%
\Pi_{\downarrow_{z}}\right\rangle _{w}+\left\langle \Pi_{R}\Pi_{\downarrow
_{z}}\right\rangle _{w}=1,\text{ }\left\langle \Pi_{\uparrow}\right\rangle
_{w}=0,
\end{align}
and so the electron has to be on the left while its spin as to be on the right
since it is down, and we recall that\ $\left\langle \Pi_{L}\Pi_{\downarrow
_{z}}\right\rangle _{w}=0$. Thus, the weak values show us that the particle is
seemingly separated from its spin. The quantum Cheshire cat effect has been
demonstrated experimentally (see, [15]).

The Hardy paradox involves two entangled particles---such as an electron and a
positron [7]. In Hardy's setup, the particles travel through a double
Mach-Zehnder interferometer, with their paths designed to interfere in such a
way that simultaneous detection of both particles (even though their
interaction would annihilate them in classical terms) can still be observed
under certain conditions. Suppose we have an electron with two position states
(two arms of the interferometer) $\left\vert L_{e}\right\rangle $ and
$\left\vert R_{e}\right\rangle $ and a positron with position states
$\left\vert L_{p}\right\rangle $ and $\left\vert R_{p}\right\rangle $ so that
$\left\vert R_{e}\right\rangle $ and $\left\vert L_{p}\right\rangle $ are
close enough to each other such that if the electron goes through its right
arm, $\left\vert R_{e}\right\rangle ,$ and the position goes through its left
arm, $\left\vert L_{p}\right\rangle $, then the particles will annihilate.

We then prepare the initial state of the system in the form%
\begin{equation}
\left\vert \psi\right\rangle =\left(  \left\vert L_{p}\right\rangle
+\left\vert R_{p}\right\rangle \right)  \left(  \left\vert L_{e}\right\rangle
+\left\vert R_{e}\right\rangle \right)  =\left\vert L_{p}\right\rangle
\left\vert L_{e}\right\rangle +\left\vert L_{p}\right\rangle \left\vert
R_{e}\right\rangle +\left\vert R_{p}\right\rangle \left\vert L_{e}%
\right\rangle +\left\vert R_{p}\right\rangle \left\vert R_{e}\right\rangle
.\label{h1}%
\end{equation}
Assuming that no annihilation has been taking place,\ we omit the state
$\left\vert L_{p}\right\rangle \left\vert R_{e}\right\rangle ,$ and our pre-
selected state is then%
\begin{equation}
\left\vert \Psi\right\rangle =\left\vert L_{p}\right\rangle \left\vert
L_{e}\right\rangle +\left\vert R_{p}\right\rangle \left\vert L_{e}%
\right\rangle +\left\vert R_{p}\right\rangle \left\vert R_{e}\right\rangle
\label{h2}%
\end{equation}
which can be written as%
\begin{equation}
\left\vert \Psi\right\rangle =\left(  \left\vert L_{p}\right\rangle
+\left\vert R_{p}\right\rangle \right)  \left\vert L_{e}\right\rangle
+\left\vert R_{p}\right\rangle \left\vert R_{e}\right\rangle \label{h3}%
\end{equation}
or%
\begin{equation}
\left\vert \Psi\right\rangle =\left\vert L_{p}\right\rangle \left\vert
L_{e}\right\rangle +\left\vert R_{p}\right\rangle \left(  \left\vert
L_{e}\right\rangle +\left\vert R_{e}\right\rangle \right)  .\label{h4}%
\end{equation}
We post-select the system into the state%
\begin{equation}
\left\vert \phi\right\rangle =\left(  \left\vert L_{p}\right\rangle
-\left\vert R_{p}\right\rangle \right)  \left(  \left\vert L_{e}\right\rangle
-\left\vert R_{e}\right\rangle \right)  .\label{postH1}%
\end{equation}
Now, since we post-select the above state, notice that $\left\vert
L_{p}\right\rangle -\left\vert R_{p}\right\rangle $ is orthogonal to
$\left\vert L_{p}\right\rangle +\left\vert R_{p}\right\rangle $ and so
following the form of the pre- selected state (\ref{h3}), the electron and the
positron has be on their right arm. Furthermore, notice that in the post-
selection we also have the electron state $\left\vert L_{e}\right\rangle
-\left\vert R_{e}\right\rangle $ which is orthogonal to the pre-selected state
$\left\vert L_{e}\right\rangle +\left\vert R_{e}\right\rangle $ appearing in
the form (\ref{h4}), implying that both the electron and the positron has be
on their left arm. So, we have found that the states $\left\vert
R_{p}\right\rangle \left\vert R_{e}\right\rangle $ and $\left\vert
L_{p}\right\rangle \left\vert L_{e}\right\rangle $ can both been occured. The
seeming paradox is seen through weak values, as suggested in [11],%

\begin{equation}
\left\langle \Pi_{L_{e}}\right\rangle _{w}=0,\text{ }\left\langle \Pi_{R_{e}%
}\right\rangle _{w}=1,\text{ }\left\langle \Pi_{L_{p}}\right\rangle
_{w}=1,\text{ }\left\langle \Pi_{R_{p}}\right\rangle _{w}=0.\label{Hw1}%
\end{equation}
Surprisingly, the electron and the positron are in the forbidden arms.
However, if we ask whether the electron went through the right arm AND the
positron went through the left arm, the answer is \textit{no}, which can be
seen through the weak value%
\begin{equation}
\left\langle \Pi_{R_{e}}\Pi_{L_{p}}\right\rangle _{w}=0.\label{Hw2}%
\end{equation}
We note that the Hardy paradox was realized in a recent experimental setup [16].

Let us now show how the above two experiments are the same at their core. We
consider the following visual scheme:

{\includegraphics[
height=2.2in,
width=2.3in
]{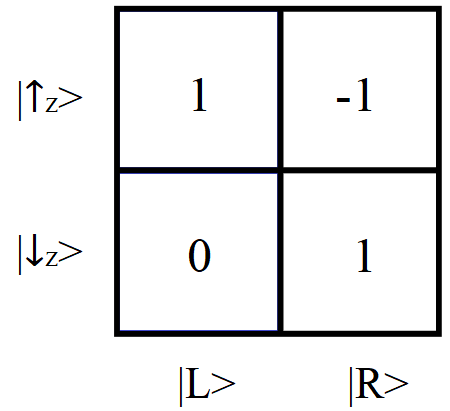}}

Figure 2. The scheme shows the equivalency between the quantum
Cheshire cat experiment that takes the same form as in the case of Hardy's paradox.

Following Figure 2, as the sum of components of the first column equals to 1
while the sum of components of the second column equals 0, we say that the
particle is on the left harm. However, as the sum of components of the first
row equals to 0 while the sum of components of the second row equals 1, the
spin of the particle is $1$. We have\ thus established a visual representation
of the\ quantum Cheshire cat experiment. Following the given weak reality with
the negative values in the scheme, we have a system of two particles in the
states $\left\vert \uparrow_{z},L\right\rangle $ and $\left\vert
\downarrow_{z},R\right\rangle $ and a \textit{negative} \textit{particle} in
the state $\left\vert \uparrow_{z},R\right\rangle $. This system of three
particles only appears in the weak reality of quantum mechanics.

For the quantum Cheshire cat effect, we interpret particle $I$ as the spin\ of
the particle in some direction while particle $II$ is the position state of
the same particle. \ We now show that when particle $I$ represents an electron
and particle $II$ represents a positron in the system, we immediately obtain
Hardy's paradox. Consider the zero and one states as being on the left and
right arms for both the electron and positron in the system. Then, we see that
while the electron goes through the right arm and the positron goes through
the left arm, the particles still interact through a negative particle. Thus,
in the weak reality, Hardy's experiment is described by three particles: a
particle goes on the left arms of the electron and positron, a particle goes
through the right arms of the electron and the positron, and a negative
particle that goes through the left arm of the electron and the right arm of
the positron. We have thus established an equivalency between the quantum
Cheshire cat experiment and Hardy's paradox.

\bigskip We can also describe the Hardy experiment by defining an operator $O
$ that represents the overlapping between the electron $(e)$ and positron
$(p)$, and $NO$ for no such overlapping, and for the positron/electron we
write $+/-$ . In particular, we define $N_{O}^{+/-}=\left\vert O\right\rangle
_{p/e}\left\langle O\right\vert _{p/e},NO_{O}^{+/-}=\left\vert NO\right\rangle
_{p/e}\left\langle NO\right\vert _{p/e},$ $N_{O,O}^{+,-}=N_{O}^{+}N_{O}%
^{-},N_{NO,NO}^{+,-}=N_{NO}^{+}N_{NO}^{-},N_{O,NO}^{+,-}=NO_{O}^{+}NO_{NO}%
^{-}, $ and $N_{NO,O}^{+,-}=NO_{NO}^{+}NO_{O}^{-}.$ Then, the Hardy paradox
leads to%
\begin{align}
\left\langle N_{NO,O}^{+,-}\right\rangle _{w}  & =+1,\left\langle
N_{NO,NO}^{+,-}\right\rangle _{w}=-1,\label{Hardy1}\\
\left\langle N_{O,O}^{+,-}\right\rangle _{w}  & =0,\text{ }\left\langle
N_{O,NO}^{+,-}\right\rangle _{w}=+1,\nonumber
\end{align}
similar to Figure 2.
In Figure 2, each value $+1$/$-1$ implies a positive/negative pair of
particles that comes from the weak values of the product of projection
operators of each quantum degree of freedom. A key element in the proposed
scheme is that from pairs, we can deduce about each particle separately, but
not vice versa. This implies that the pairs are more fundamental in the
proposed model, and in general, when dealing with groups of $K$ particles in
the form of tensor representation as will be shown in (\ref{amT1}), such
groups of particles is the fundamental structure of the model. The reason such
pairs are more fundamental is that by summing up the amplitudes for the pairs
we deduce the values for each of the particles, separately. In our case, we
have $6$ particles -- $4$ particles and $2$ \textit{negative} particles, since
each $+1/-1$ is a pair of particles. For each of the particles, we then have
$\left\langle N_{O}^{-}\right\rangle _{w}=+1,\left\langle N_{O}^{+}%
\right\rangle =+1,\left\langle N_{NO}^{-}\right\rangle =0=1-1,$ and
$\left\langle N_{NO}^{+}\right\rangle =0=1-1,$ implying on four particles and
two \textit{negative} particles. In the visual representations of each
(physical) particles, we can deduce the pairs, as follows:

{\includegraphics[
height=2.3in,
width=2.3in
]{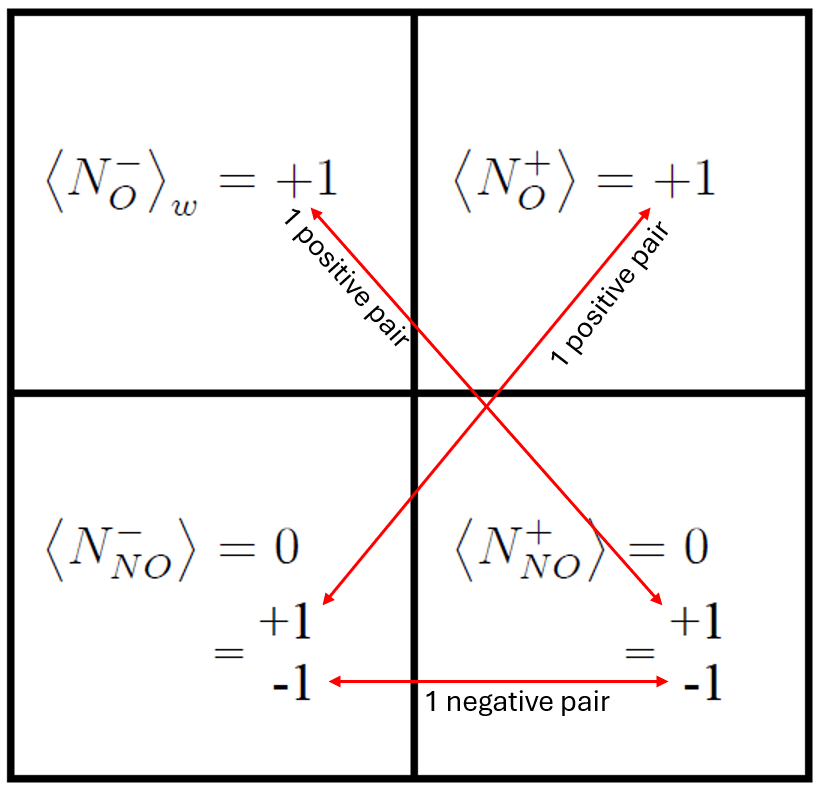}}

Figure 3. A\ single particle scheme for the Hardy paradox, where each arrow
represents a pair of particles, summing up to $6$ particles.

\bigskip

The proposed framework is general and can be extended into quantum systems
with an arbitrary number of eigenstates. For instance, we can extend both the
quantum Cheshire cat effect and Hardy's paradox using the following setup:
{\includegraphics[
height=2.2in,
width=2.3in
]{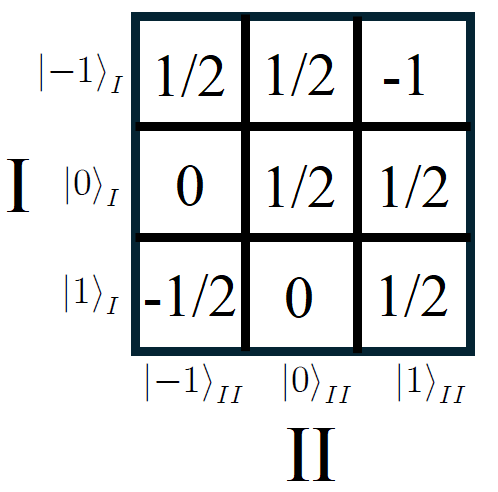}}

Figure 4. A quantum system with three eigenstates for each realized
quantum particle.

For the quantum Cheshire cat experiment dealing with spin as the property that
is separated from the particle, we extend it to a particle with spin $1$ and
for three possible boxes/arms, e.g., left-center-right. Then, the experiment
can be done, for example, for a W-boson between its position and the position
of its spin. However, in that case, Hardy's paradox cannot hold for W-bosons,
which have spin $1$, since a boson is the antiparticle of itself and, thus,
cannot be annihilated as proposed in the original Hardy experiment. We can,
however, consider other spin $1$ particles for the experiment, such as the
$W^{\pm}$ bosons where $W^{-}$ is the antiparticle of $W^{+}$. Following the
proposed setup, we can extend the quantum Cheshire cat such that the particle
goes through a single path while the spin goes through the other $k-1$ paths.
Note that the dimensions of the square table dictate whether we are dealing
with bosons or fermions. In the case of an even(odd) dimension, we are dealing
with bosons(fermions). In the case of odd dimensions $N\geq3,$ we still have
to define the particle for obtaining the Hardy paradox, e.g., in the case of a
Majorana fermion, we have a particle that is its own antiparticle, so Hardy's
effect does not take place. The equivalency between the quantum Cheshire cat
and Hardy's paradox only holds onto the mathematical description given by the
proposed universal framework. However, by providing an interpretation to each
set of quantum states, i.e., assigning specific properties, we can introduce a
totally new quantum effect.

We introduce a class of experiments that generalize both the quantum Cheshire
cat experiment and Hardy experiment, including the entire possible experiments
that can be derived by the scheme of Figure 4, using the parametrized scheme:

{\includegraphics[
height=2.6in,
width=2.8in
]{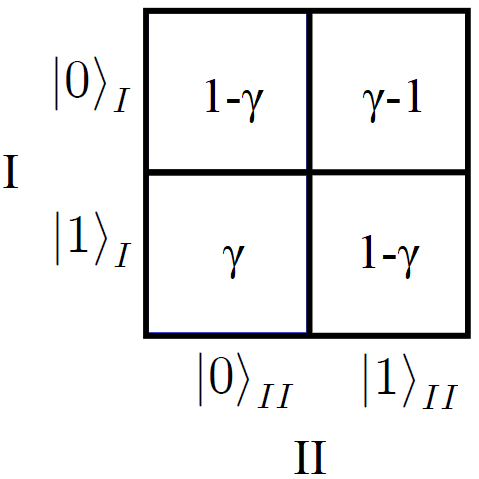}}

Figure 5. A\ quantum system of qubits with very small coupling quantified by
the parameter $0<\gamma<<1$.

\bigskip

Following Figure 5, instead of assuming annihilation in the case of
interaction between the electron and positron, the interaction gives rise to a
relative phase in the system quantified by the parameter $\gamma.$ An
extension of the proposed scheme can be made when considering more complex
setups. Let us now show how the proposed framework allows us to obtain a
system of two pairs of EPR\ states, realized by two physical particles $1$ and
$2,$ with realizations of $\left(  \left\vert x\right\rangle _{I}^{\left(
j\right)  },\left\vert x\right\rangle _{II}^{\left(  j\right)  }\right)
,j=1,2.$ Each EPR\ state is given by a square packed with four equal squares.
Now, suppose that the pair of squares are tangent, such that each of the
given\ systems is in the system $\left\vert 1\right\rangle _{I}^{\left(
j\right)  }\left\vert 1\right\rangle _{II}^{\left(  j\right)  },$ and there is
a possibility of tunneling between the boxes $\left\vert 11\right\rangle
^{\left(  1\right)  }$ and $\left\vert 10\right\rangle ^{\left(  2\right)  },$
see Figure 6 below.
{\includegraphics[
height=2.1in,
width=3.4in
]{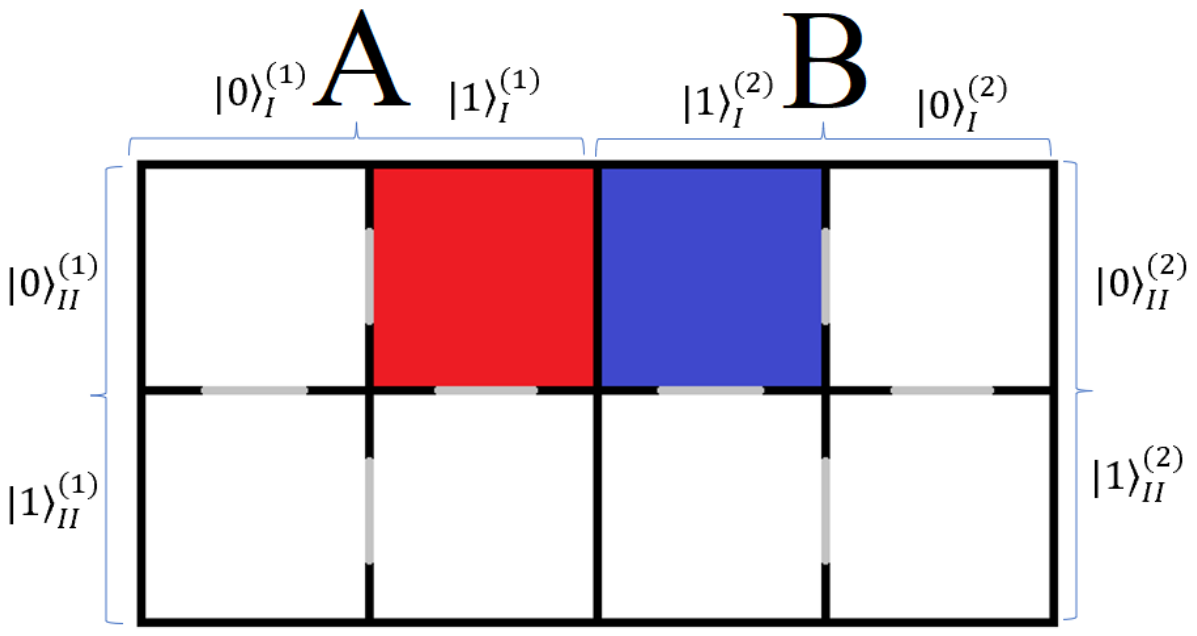}}

Figure 6. The $4-$qubits multiwise interaction, given by local interactions of
the red and blow boxes, where the position of the physical particle $1$ is in
the red box, while the position of the physical particle $2$ is in the blue box.

The Hamiltonian of the realized multi-interaction is given by
\begin{equation}
H=\varepsilon\left\vert 0\right\rangle _{II}^{\left(  1\right)  }\left\langle
0\right\vert _{II}^{\left(  1\right)  }\otimes\left\vert 1\right\rangle
_{I}^{\left(  1\right)  }\left\langle 1\right\vert _{I}^{\left(  1\right)
}\otimes\left\vert 0\right\rangle _{II}^{\left(  2\right)  }\left\langle
0\right\vert _{II}^{\left(  2\right)  }\otimes\left\vert 1\right\rangle
_{I}^{\left(  2\right)  }\left\langle 1\right\vert _{I}^{\left(  2\right)
},\label{He11}%
\end{equation}
and we\ start with the product state of the product of EPR\ states $\left\vert
\Psi_{EPR-EPR}\left(  t=0\right)  \right\rangle =\left\vert EPR\right\rangle
_{1}\otimes\left\vert EPR\right\rangle _{2}$ where $\left\vert
EPR\right\rangle _{i}=\left\vert 1\right\rangle _{I}^{\left(  i\right)
}\otimes\left\vert 0\right\rangle _{II}^{\left(  i\right)  }-\left\vert
0\right\rangle _{I}^{\left(  i\right)  }\otimes\left\vert 1\right\rangle
_{II}^{\left(  i\right)  }.$

Then, at time $t,$ the state is given by
\begin{align}
\left\vert \Psi_{EPR-EPR}\left(  t\right)  \right\rangle  & =e^{-iH_{1}%
t}\left\vert EPR\right\rangle _{1}\otimes e^{-iH_{2}t}\left\vert
EPR\right\rangle _{2}\label{psit1}\\
& =\frac{1}{2}\left(  \left\vert 1\right\rangle _{I}^{\left(  1\right)
}\left\vert 0\right\rangle _{II}^{\left(  1\right)  }e^{-i\varepsilon
t}-\left\vert 0\right\rangle _{I}^{\left(  1\right)  }\left\vert
1\right\rangle _{II}^{\left(  1\right)  }\right) \nonumber\\
& \otimes\left(  \left\vert 1\right\rangle _{I}^{\left(  2\right)  }\left\vert
0\right\rangle _{II}^{\left(  2\right)  }e^{-i\varepsilon t}-\left\vert
0\right\rangle _{I}^{\left(  2\right)  }\left\vert 1\right\rangle
_{II}^{\left(  2\right)  }\right)  .\nonumber
\end{align}
The $4-$qubit interaction is governed by the coupling constant $\varepsilon$,
where any change in one EPR state immediately changes the relative phase of
the other EPR state. This provides a unique and new type of interaction
between more than two qubits per interaction.

We can generalize the given procedure into $n>2$ (multiwise) interacted
qubits. Starting with a set of three squares that are geometrically defined as
a cube, we have the following interaction Hamiltonian%
\begin{align}
H  & =\varepsilon\left\vert 0\right\rangle _{II}^{\left(  1\right)
}\left\langle 0\right\vert _{II}^{\left(  1\right)  }\otimes\left\vert
1\right\rangle _{I}^{\left(  1\right)  }\left\langle 1\right\vert
_{I}^{\left(  1\right)  }\otimes\left\vert 0\right\rangle _{II}^{\left(
2\right)  }\left\langle 0\right\vert _{II}^{\left(  2\right)  }\label{HH11}\\
& \otimes\left\vert 1\right\rangle _{I}^{\left(  2\right)  }\left\langle
1\right\vert _{I}^{\left(  2\right)  }\otimes\left\vert 0\right\rangle
_{II}^{\left(  3\right)  }\left\langle 0\right\vert _{II}^{\left(  3\right)
}\otimes\left\vert 1\right\rangle _{I}^{\left(  3\right)  }\left\langle
1\right\vert _{I}^{\left(  3\right)  }.\nonumber
\end{align}
Then, as before, we start with the product state of the EPR\ states which now
considers three EPR pairs, and we obtain%
\begin{align}
& \left\vert \Psi_{EPR-EPR-EPR}\left(  t\right)  \right\rangle \label{E111}\\
& =e^{-iH_{1}t}\left\vert EPR\right\rangle _{1}\otimes e^{-iH_{2}t}\left\vert
EPR\right\rangle _{2}\otimes e^{-iH_{3}t}\left\vert EPR\right\rangle
_{3}\nonumber\\
& =\frac{1}{2}\bigotimes\limits_{j=1}^{3}\left(  \left\vert 1\right\rangle
_{I}^{\left(  j\right)  }\left\vert 0\right\rangle _{II}^{\left(  j\right)
}e^{-i\varepsilon t}-\left\vert 0\right\rangle _{I}^{\left(  j\right)
}\left\vert 1\right\rangle _{II}^{\left(  j\right)  }\right)  .\nonumber
\end{align}

Thus, while in nature, interactions come only in pairs, it does not forbid
multi-interactions of groups containing $n>2$ qubits.

We can further consider a six multi-interacted qubits interaction, but with
pairwise coupling, followed by the system given in Figure 7.

\bigskip
{\includegraphics[
height=1.8in,
width=3.4in
]{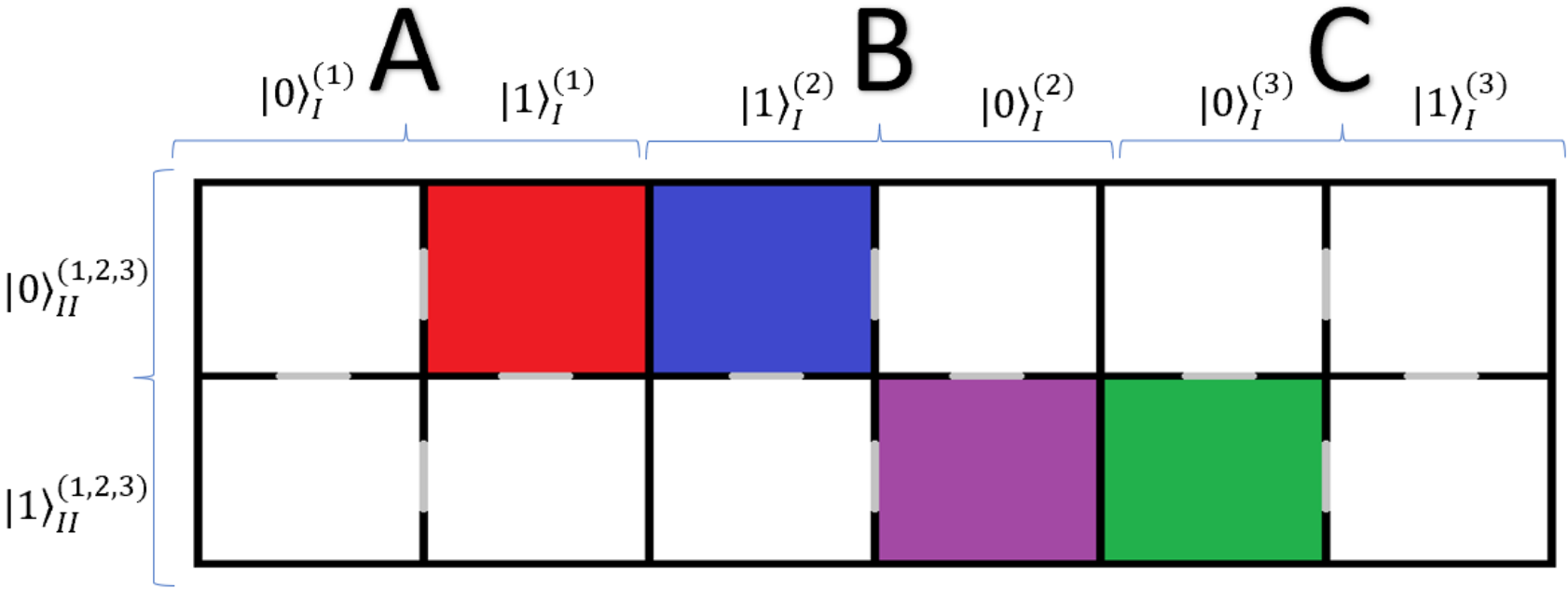}}

Figure 7. A system of $6$ realized qubits that construct $3$ EPR states.

\bigskip

In Figure 7, we illustrate three EPR coupled EPR states where the interaction
Hamiltonian is given by%
\begin{align}
H  & =\varepsilon_{1}\left\vert 0\right\rangle _{II}^{\left(  1\right)
}\left\langle 0\right\vert _{II}^{\left(  1\right)  }\otimes\left\vert
1\right\rangle _{I}^{\left(  1\right)  }\left\langle 1\right\vert
_{I}^{\left(  1\right)  }\otimes\left\vert 0\right\rangle _{II}^{\left(
2\right)  }\left\langle 0\right\vert _{II}^{\left(  2\right)  }\otimes
\left\vert 1\right\rangle _{I}^{\left(  2\right)  }\left\langle 1\right\vert
_{I}^{\left(  2\right)  }\label{Hamm1}\\
& +\varepsilon_{2}\left\vert 1\right\rangle _{II}^{\left(  1\right)
}\left\langle 1\right\vert _{II}^{\left(  1\right)  }\otimes\left\vert
0\right\rangle _{I}^{\left(  1\right)  }\left\langle 0\right\vert
_{I}^{\left(  1\right)  }\otimes\left\vert 1\right\rangle _{II}^{\left(
2\right)  }\left\langle 1\right\vert _{II}^{\left(  2\right)  }\otimes
\left\vert 0\right\rangle _{I}^{\left(  2\right)  }\left\langle 0\right\vert
_{I}^{\left(  2\right)  }\nonumber
\end{align}

Then, as before, we start with the product state of the EPR\ states which now
considers three EPR pairs, and we obtain%
\begin{align}
& \left\vert \Psi_{EPR-EPR-EPR}\left(  t\right)  \right\rangle \label{Hamm2}\\
& =e^{-iH_{1}t}\left\vert EPR\right\rangle _{1}\otimes e^{-iH_{2}t}\left\vert
EPR\right\rangle _{2}\otimes e^{-iH_{3}t}\left\vert EPR\right\rangle
_{3}\nonumber\\
& =\left(  \left\vert 1\right\rangle _{I}^{\left(  j\right)  }\left\vert
0\right\rangle _{II}^{\left(  j\right)  }e^{-i\varepsilon_{1}t}-\left\vert
0\right\rangle _{I}^{\left(  j\right)  }\left\vert 1\right\rangle
_{II}^{\left(  j\right)  }\right)  \otimes\left(  \left\vert 1\right\rangle
_{I}^{\left(  j\right)  }\left\vert 0\right\rangle _{II}^{\left(  j\right)
}e^{-i\left(  \varepsilon_{1}-\varepsilon_{2}\right)  t}-\left\vert
0\right\rangle _{I}^{\left(  j\right)  }\left\vert 1\right\rangle
_{II}^{\left(  j\right)  }\right) \nonumber\\
& \otimes\left(  \left\vert 1\right\rangle _{I}^{\left(  j\right)  }\left\vert
0\right\rangle _{II}^{\left(  j\right)  }-\left\vert 0\right\rangle
_{I}^{\left(  j\right)  }\left\vert 1\right\rangle _{II}^{\left(  j\right)
}e^{-i\varepsilon_{2}t}\right)  .\nonumber
\end{align}
in this way, we can get an effect where a change in particles 1 and 3, in
their relative phase, will not give a change in particle 2.

\bigskip

\subsection{\bigskip Extensions to tensors}

We can establish a universal framework for each quantum experiment that is
based on both weak and strong realities by introducing a tensor of dimensions
$d_{1}\times d_{2}\times...\times d_{N},$ with the components%
\begin{equation}
a_{i_{1}^{\left[  d_{1}\right]  }i_{2}^{\left[  d_{2}\right]  }...i_{N}%
^{\left[  d_{N}\right]  }}=\frac{\left\langle \phi\left\vert
%TCIMACRO{\tprod _{j=1}^{N}}%
%BeginExpansion
{\textstyle\prod_{j=1}^{N}}
%EndExpansion
\Pi_{i_{j}^{\left[  d_{j}\right]  }}^{I^{\left(  j\right)  }}\right\vert
\psi\right\rangle }{\left\langle \phi|\psi\right\rangle }\label{amT1}%
\end{equation}
where $i_{j}^{\left[  d_{1}\right]  }$ is an index that runs from
$i_{1}^{\left[  d_{1}\right]  }=1$ to $i_{1}^{\left[  d_{1}\right]  }=d_{1},$
and $I^{\left(  1\right)  }=I,$ $I^{\left(  2\right)  }=II,$ $I^{\left(
3\right)  }=III,$ etc., which describe a quantum system of $N$ sub-systems,
with the projection operators%
\begin{equation}
\Pi_{i_{j}^{\left[  d_{j}\right]  }}^{I^{\left(  j\right)  }}=\left\vert
i_{j}^{\left[  d_{j}\right]  }\right\rangle _{I^{\left(  j\right)  }%
}\left\langle i_{j}^{\left[  d_{j}\right]  }\right\vert _{I^{\left(  j\right)
}}.\label{PO11}%
\end{equation}
We note that the tensor does not have to be of the same number of components
for each dimension, e.g., not having a square table. This implies that the
number of states for each property can be different, e.g., in a quantum
Cheshire cat experiment for a W-boson and its spin where spin can be in one of
three states, but the position can be in the left or right boxes. The general
framework (\ref{PO11}) allows us to generate novel quantum effects. For
example, we can extend the quantum Cheshire cat experiment by considering the
sequence $I^{\left(  1\right)  },I^{\left(  2\right)  },...,I^{\left(
k\right)  },I^{\left(  k+1\right)  },...,I^{\left(  N\right)  }$ as
representing a molecule consists of $k$ nuclei and $N-k$ electrons. Then,
under the given pre- and post- selected states of the molecule, we can
establish an effect in which the electrons are bounded much like in the
standard molecular structure, but, around empty space. We call these
structures \textit{hollow molecules} as an extension of the hollow atom
proposed in [17].

We can consider the GHZ state for a quantum system of three particles with
eigenvalues $0,1,2.$ The GHZ state is then $\left\vert GHZ\right\rangle
=\left(  \left\vert 000\right\rangle +\left\vert 222\right\rangle \right)
/\sqrt{2}.$ Then, the standard correlations can be obtained via the formula%
\begin{equation}
a_{i_{1}^{\left[  d_{1}\right]  }i_{2}^{\left[  d_{2}\right]  }i_{3}^{\left[
d_{3}\right]  }}=\left\langle \psi\left\vert \Pi_{i_{1}^{\left[  d_{1}\right]
}}^{I^{\left(  1\right)  }}\Pi_{i_{2}^{\left[  d_{2}\right]  }}^{I^{\left(
2\right)  }}\Pi_{i_{3}^{\left[  d_{3}\right]  }}^{I^{\left(  3\right)  }%
}\right\vert \psi\right\rangle \label{aa01}%
\end{equation}
\ that gives us $a_{000}=\left\langle GHZ\left\vert \Pi_{0}^{I}\Pi_{0}^{II}%
\Pi_{0}^{III}\right\vert GHZ\right\rangle =1/2,$ $a_{222}=\left\langle
GHZ\left\vert \Pi_{2}^{I}\Pi_{2}^{II}\Pi_{2}^{III}\right\vert GHZ\right\rangle
=1/2,$ and zero otherwise. Now, following the proposed scheme, we can extend
it into a more general quantum coupling, followed by (\ref{amT1}). For this,
we consider the pre- selected state $\left\vert \psi\right\rangle =\left(
\left\vert 000\right\rangle +\left\vert 111\right\rangle +\left\vert
222\right\rangle \right)  /\sqrt{3}$ and post- selected state $\left\vert
\psi\right\rangle =\left(  \left\vert 000\right\rangle +\left\vert
111\right\rangle -\left\vert 222\right\rangle \right)  /\sqrt{3}.$ This then
leads to the values%
\begin{equation}
a_{000}=1,\text{ }a_{111}=1,\text{ }a_{222}=-1,\text{ }a_{ijk}=0,\text{
\thinspace}\forall\,i\neq j\neq k,\label{aa02}%
\end{equation}
leading to a total number of $1,$ as expected. Unlike the GHZ state, which is
described by a ket state in the form of a (normalized) sum of $\left\vert
jjj\right\rangle $ states, here, the coupling is only deduced through the
components $a_{ijk}.$

\bigskip
{\includegraphics[
height=3in,
width=3in
]{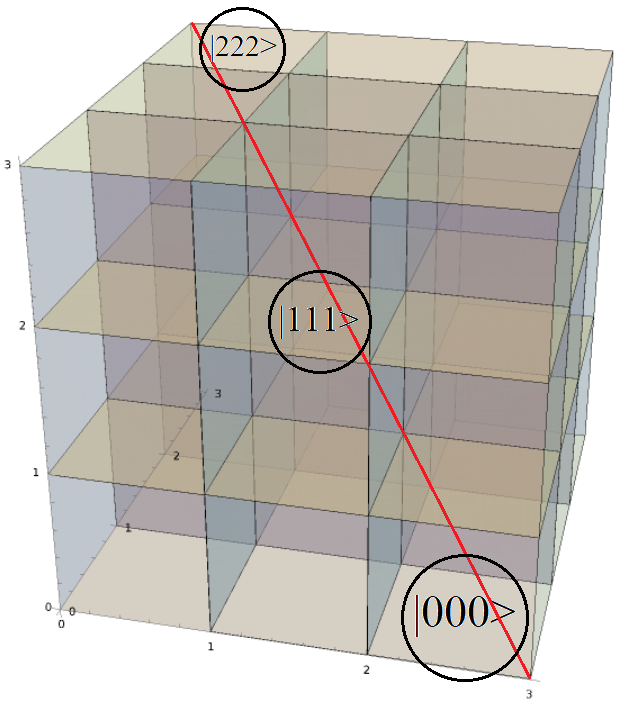}}

Figure 8. a visual representation of the $3-$particle system with three
eigenvalues. The diagonal represents the states involved in the entangled structure.

\bigskip

\bigskip

Following Figure 8, we see that in the visual representation, there is a
special role for the diagonals and edges of the cubes. The reason is that if
we would consider other components of our cube, it will not necessarily imply
coupling, e.g., $\left\vert 000\right\rangle +\left\vert 001\right\rangle
=\left\vert 00\right\rangle \left(  \left\vert 0\right\rangle +\left\vert
1\right\rangle \right)  .$ For future research, we proposed to explore how
such diagonals and edges that can also appear on the hypercube when
considering a system of $N$ particles lead to maximum correlations between the
entangled particles.

\subsubsection{Multi-interacted GHZ\ states}

Using the proposed visual representation, we now realize unique interacted
particles that go beyond the standard pairwise entangled systems that appear
in nature.\ Let us consider a pair of cubes, where each of the cubes is
occupied by a single particle that can be in one of the squares given by half
of the surface of the cube, as given in the following Figure:

{\includegraphics[
height=2.1in,
width=3.4in
]{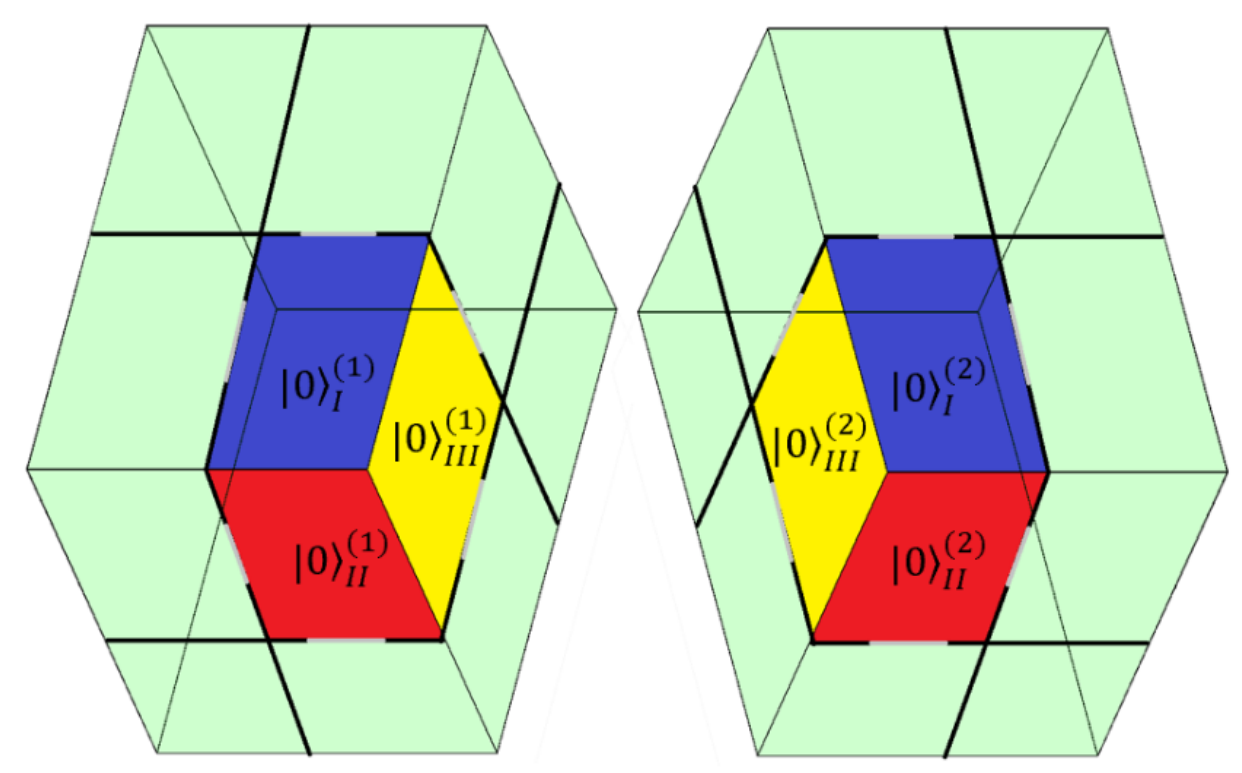}}

Figure 9. A\ pair of cubes, each of them containing a single physical state,
occupy 12 packed squares in 3 squares of the cube's surface. Blue, Red, and
Yellow (BRY) describe the $I,II,III$ realized qubits for each of the cubes.

\bigskip

By having\ tunneling for each of the cubes that corresponds to $\sigma
_{x}\otimes\sigma_{x}\otimes\sigma_{x},$ i.e., tunneling between the
$\left\vert 0\right\rangle _{I/II/III}^{\left(  1,2\right)  }$ state and its
neighbors on the surface, we produce a GHZ\ state for each of the cubes, so we
start with a product GHZ state $\left\vert GHZ\right\rangle _{1}%
\otimes\left\vert GHZ\right\rangle _{2}.$ Then, suppose we have an interaction
Hamiltonian, as in the EPR\ case%
\begin{align}
H_{int}  & =\phi\left\vert 0\right\rangle _{III}^{\left(  1\right)
}\left\langle 0\right\vert _{III}^{\left(  1\right)  }\otimes\left\vert
0\right\rangle _{II}^{\left(  1\right)  }\left\langle 0\right\vert
_{II}^{\left(  1\right)  }\otimes\left\vert 0\right\rangle _{I}^{\left(
1\right)  }\left\langle 0\right\vert _{I}^{\left(  1\right)  }\label{HintGHZ1}%
\\
& \otimes\left\vert 0\right\rangle _{III}^{\left(  2\right)  }\left\langle
0\right\vert _{III}^{\left(  2\right)  }\otimes\left\vert 0\right\rangle
_{II}^{\left(  2\right)  }\left\langle 0\right\vert _{II}^{\left(  2\right)
}\otimes\left\vert 0\right\rangle _{I}^{\left(  2\right)  }\left\langle
0\right\vert _{I}^{\left(  2\right)  }\nonumber
\end{align}
which changes the potential energy at each of the BRY sub-cubes by a constant
$\phi>0,$ depending on how close the cubes are. Then, when they are close
enough, we produce the state%
\begin{align}
& \left\vert \Psi_{GHZ-GHZ}\left(  t\right)  \right\rangle \label{GHZ2}\\
& =\frac{1}{2}\left(  \left\vert 0\right\rangle _{I}^{\left(  1\right)
}\left\vert 0\right\rangle _{II}^{\left(  1\right)  }\left\vert 0\right\rangle
_{III}^{\left(  1\right)  }e^{-i\phi t/\hbar}+\left\vert 1\right\rangle
_{I}^{\left(  1\right)  }\left\vert 1\right\rangle _{II}^{\left(  1\right)
}\left\vert 1\right\rangle _{III}^{\left(  1\right)  }\right) \nonumber\\
& \otimes\left(  \left\vert 0\right\rangle _{I}^{\left(  2\right)  }\left\vert
0\right\rangle _{II}^{\left(  2\right)  }\left\vert 0\right\rangle
_{III}^{\left(  2\right)  }e^{-i\phi t/\hbar}+\left\vert 1\right\rangle
_{I}^{\left(  2\right)  }\left\vert 1\right\rangle _{II}^{\left(  2\right)
}\left\vert 1\right\rangle _{III}^{\left(  2\right)  }\right)  .\nonumber
\end{align}
which realizes $6$ coupled qubits in pairs of GHZ states.

Suppose now, we put another cube at the $\left\vert 1\right\rangle
_{I}^{\left(  2\right)  }\left\vert 1\right\rangle _{II}^{\left(  2\right)
}\left\vert 1\right\rangle _{III}^{\left(  3\right)  }$ sub-cube, with the
same interaction as (\ref{HintGHZ1}) with a parameter $\beta=\phi+\varepsilon
$, so we have%
\begin{align}
& \left\vert \Psi_{GHZ-GHZ}\left(  t\right)  \right\rangle \label{PsiGHZ11}\\
& =\frac{1}{2\sqrt{2}}\left(  \left\vert 0\right\rangle _{I}^{\left(
1\right)  }\left\vert 0\right\rangle _{II}^{\left(  1\right)  }\left\vert
0\right\rangle _{III}^{\left(  1\right)  }e^{-i\phi t/\hbar}+\left\vert
1\right\rangle _{I}^{\left(  1\right)  }\left\vert 1\right\rangle
_{II}^{\left(  1\right)  }\left\vert 1\right\rangle _{III}^{\left(  1\right)
}\right) \nonumber\\
& \otimes\left(  \left\vert 0\right\rangle _{I}^{\left(  2\right)  }\left\vert
0\right\rangle _{II}^{\left(  2\right)  }\left\vert 0\right\rangle
_{III}^{\left(  2\right)  }e^{i\varepsilon t/\hbar}+\left\vert 1\right\rangle
_{I}^{\left(  2\right)  }\left\vert 1\right\rangle _{II}^{\left(  2\right)
}\left\vert 1\right\rangle _{III}^{\left(  2\right)  }\right) \nonumber\\
& \otimes\left(  \left\vert 0\right\rangle _{I}^{\left(  3\right)  }\left\vert
0\right\rangle _{II}^{\left(  3\right)  }\left\vert 0\right\rangle
_{III}^{\left(  3\right)  }+\left\vert 1\right\rangle _{I}^{\left(  3\right)
}\left\vert 1\right\rangle _{II}^{\left(  3\right)  }\left\vert 1\right\rangle
_{III}^{\left(  3\right)  }e^{-i\beta t/\hbar}\right)  .\nonumber
\end{align}
In this case, we see that the parameter $\varepsilon$ controls the coupling
for the $9-$interacted qubits. When $\varepsilon\neq0,$ we have coupling
between the three\ states. However, when $\varepsilon=0,$ the second cube has
a constant relative phase of zero, which gives the usual GHZ state,
$\left\vert GHZ\right\rangle _{2}.$

\section{Discussion}

We have established a visual framework for exploring entangled quantum
systems. The proposed framework provides a way to explore the foundations of
different quantum effects to gain a deeper insight into the foundations of
entanglement in quantum mechanics and the emergence of quantum phenomena. We
found it interesting that the important entangled states - the Bell and GHZ
states- are described using the proposed (squares)cubes followed by the
diagonal states. We conjecture that this also applies in the case of
$N>>2$\ dimensions, i.e., the interesting entangled states emerge from
considering the diagonal states of such hypercubes. Using the proposed
framework, we have established\ a way to detect equivalencies between
different experiments. For example, we have shown the equivalency between the
quantum Cheshire cat and the Hardy experiments. The proposed framework allowed
us to establish a pathway for deriving novel quantum effects. The proposed
framework also allowed us to realize quantum interacted systems that cannot be
obtained using physical qubits. In particular, we have shown how to realize
quantum systems that include coupling between $m>2$ qubits simultaneously,
i.e., having multi-wise interactions rather than pairwise interactions that
are the only ones to appear in nature. This can potentially have deep
implications in studying the structure of quantum mechanics, including
potential implications in quantum computing, in particular in quantum
simulation algorithms. We intend to explore such a unique type of coupling in
future research. Moreover, for future research, we intend to use the proposed
framework to gain insights into additional equivalencies between quantum
phenomena of many particles in order to achieve novel quantum phenomena and
novel insights into the nature of entanglement and nonlocality. Following the
proposed model, any tensor whose components are summed up to $1$ applies to a
different experimental setup. We intend to explore the detection of novel
effects using AI tools for detecting such tensors for given sets of assigned
properties of the particles, e.g., positions, spin, circular polarization, and
angular momentum,\ that lead to potentially interesting quantum effects.

\end{document}